\documentclass{nature}

\usepackage{amsmath}
\usepackage{url}
 \usepackage[]{graphicx}
 \usepackage{pdfpages}
 \usepackage{colortbl}
\definecolor{lightgray}{rgb}{0.95, 0.95, 0.95}
\usepackage{multirow}
\usepackage{array}
\usepackage{adjustbox}
\usepackage{tabularx}

\usepackage{longtable}

\title{The intensification of winter mid-latitude storm tracks
in the Southern Hemisphere}

\author{{\normalsize Rei Chemke$^{1}$ \& Yi Ming$^{2}$ \& Janni Yuval$^{3}$}}

\begin{document}


\maketitle
\vspace*{-0.6cm}

\begin{affiliations}
\small \item Department of Earth and Planetary Sciences,
Weizmann Institute of Science, Rehovot, Israel
\item NOAA/Geophysical Fluid Dynamics Laboratory, Princeton,
New Jersey, USA
\item Department of Earth, Atmospheric
and Planetary Sciences, MIT, Cambridge,
Massachusetts, USA
\end{affiliations}
\vspace*{-0.5cm}

\begin{abstract}
The strength of mid-latitude storm tracks shapes weather and climate
phenomena in the extra-tropics, as these storm tracks control the
daily to multi-decadal variability of precipitation, temperature and
winds. By the end of this century, winter mid-latitude storms are
projected to intensify in the Southern Hemisphere, with large consequences
over the entire extra-tropics. Therefore, it is critical to be able
to accurately assess the impacts of anthropogenic emissions on these
storms, in order to improve societal preparedness for future changes.
Here we show that current climate models severely underestimate the
intensification in mid-latitude storm-tracks in recent decades. Specifically,
the intensification obtained from reanalyses has already reached the
model-projected end of the century intensification. The biased intensification
is found to be linked to biases in the zonal flow. These results question
the ability of climate models to accurately predict the future impacts
of anthropogenic emissions in the Southern Hemisphere mid-latitudes.

\end{abstract}
\vspace*{0.6cm}
\noindent Corresponding author: Rei Chemke$^{1}$; Email: rei.chemke@weizmann.ac.il
\pagebreak

\paragraph{Main}~\\Mid-latitude storms transfer momentum, moisture and heat across different
latitudes and longitudes, thus controlling the distribution of winds,
precipitation and temperature over the extra-tropics\cite{Pfahl2012,Catto2015,Ma2017,Yau2020}.
By the end of this century, climate models project a robust strengthening
of winter mid-latitude storms in the Southern Hemisphere\cite{Yin2005,O'Gorman2010b,Wu2010,Chang2012,Harvey2014,Lehmann2014,Shaw2016a,Chang2017,Blazquez2019},
which will have large climate consequences for the entire Southern
Hemisphere extra-tropics\cite{Bengtsson2009,Yettella2017}. 

Over the 1980-2012 period, Southern Hemisphere winter cyclones were
also found to intensify in one reanalysis\cite{Reboita2015}. However,
whether this recent strengthening of winter mid-latitude storms is
part of the emerged forced response or merely part of internal climate
variability is still an open question. The answer to this question
will reveal part of the impacts of anthropogenic emissions on the
mid-latitude circulation in recent decades, which will allow policymakers
to construct more accurate adaption strategies. Another motivation
for investigating the recent changes in the mid-latitude flow comes
from previous studies who documented model biases in the climatological
(time mean) Southern Hemisphere winter circulation, including an equatorward
bias in the climatological position of the mean zonal wind\cite{Kidston2010c,Simpson2016,Bracegirdle2020},
and an underestimation of the climatological strength of high-intensity
cyclones\cite{Priestley2020}. Thus, investigating the recent changes
in the intensity of wintertime mid-latitude storms in models will
also allow us to evaluate how well climate models reproduce the trends
of mid-latitude storms over the last decades, and potentially discover
biases in the projections of storms in current climate models. We
here focus on intensity changes of the storm tracks during winter,
since during summer, the storm tracks are projected to shift poleward,
which results in minor changes in their mid-latitude mean intensity\cite{Lehmann2014}.

\begin{figure*}
\centering{}\includegraphics[width=1\columnwidth]{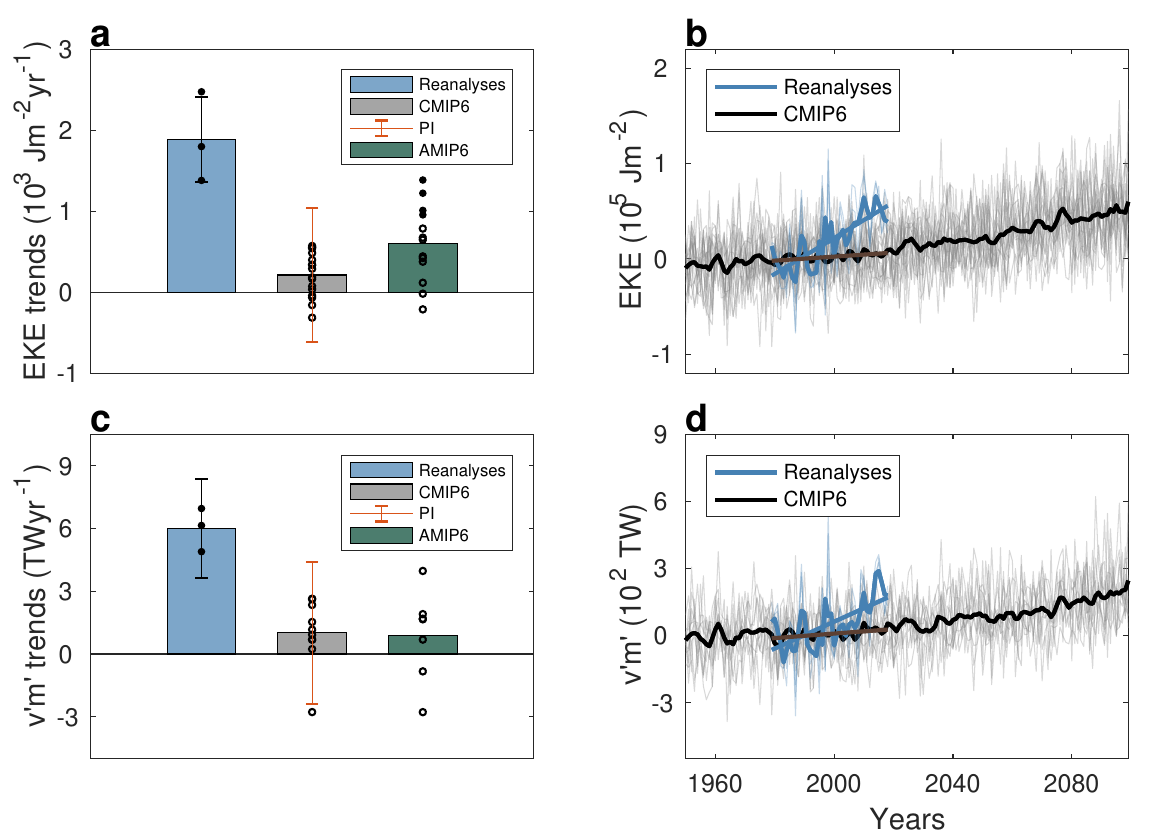}\protect\caption{\label{fig:trends}\textbf{Recent changes in Southern Hemisphere winter
mid-latitude storm tracks. }The 1979-2018 trends in \textbf{a,}~EKE
and \textbf{b,}~$v'm'$ in reanalyses mean (blue) and CMIP6 mean
(gray). The green bars show the AMIP6 mean trends over the 1979-2014
period. The black circles show the trends from the individual reanalyses/models,
where filled (open) circles show trends that are (not) statistically
significant at the $5\%$ level based on a Student's t-test. The black
bars show the $95\%$ confidence interval of the mean reanalyses trend
and red bars show the two standard deviation across all 40-year trends
from pre-industrial runs. Time evolution of \textbf{b,}~EKE and \textbf{d,}~$v'm'$,
relative to the 1980-1999 period, in reanalyses mean (blue) and CMIP6
mean (black). Thin lines show the evolution of individual reanalyses/models.
The ensemble mean time evolution is smoothed with a 3-point running
mean for plotting purposes. Blue and brown lines show the 1979-2018
linear regressions in reanalyses and CMIP6 mean, respectively. The
$v'm'$ trends are shown for $50^{\circ}{\rm {S}}$, but similar results
are evident throughout the mid-latitudes (Extended Data Fig.~2).}
\end{figure*}

\paragraph{Recent trends in Southern Hemisphere winter storm tracks}~\\We start by examining the recent changes in the intensity of Southern
Hemisphere winter (June-August; results for September-November are
shown in Supplementary Fig.~1) mid-latitude storm tracks using the
transient eddy kinetic energy (EKE, Methods\cite{O'Gorman2010b,Chang2012,Coumou2015,Chemke2020c}).
Specifically, we focus on the 40-year trends (1979-2018) of mid-latitude
EKE in 3 different reanalyses and 16 models (Fig.~\ref{fig:trends}a)
that participate in the Coupled Model Intercomparison Project Phase
6\cite{Eyring2016} (CMIP6), forced with the Historical and the SSP5-8.5
future scenario (Methods). We find that in reanalyses, the EKE have
intensified over the last four decades in a mean rate of $1.8\times10^{3}\,{\rm {Jm^{-2}yr^{-1}}}$
(blue bar; varying between $1.4\times10^{3}-2.5\times10^{3}\,{\rm {Jm^{-2}yr^{-1}}}$
across the reanalyses, black circles)\cite{Reboita2015}. In contrast,
CMIP6 models simulate a much weaker strengthening, which varies across
the models between $-315\,{\rm {Jm^{-2}yr^{-1}}}$ and $570\,{\rm {Jm^{-2}yr^{-1}}}$
(black circles), with a multi-model mean value of $210\,{\rm {Jm^{-2}yr^{-1}}}$
(gray bar). Thus, not a single model is able to capture the intensification
of the EKE in reanalyses (not even when including the uncertainty
in the mean reanalyses trend\cite{Thompson2015}; black bar). This
models-reanalyses discrepancy is not only evident over the entire
1979-2018 period, but also reanalyses show larger 10-, 20- and 30-year
trends over the 1979-2018 period (Extended Data Fig.~1).

Examining the time evolution of EKE further reveals the large differences
between reanalyses and climate models (Fig.~\ref{fig:trends}b).
First, climate models simulate a monotonic strengthening of the EKE
over the 20th and 21st centuries\cite{Yin2005,O'Gorman2010b,Wu2010,Chang2012,Harvey2014,Lehmann2014,Shaw2016a,Chang2017,Blazquez2019}.
In particular, over the last decade, the EKE in CMIP6 models has intensified
by $\sim2\%$ (with standard deviation of $\sim\pm2\%$ across the
models), relative to the 1980-1999 period, while in reanalyses the
EKE have intensified by $\sim12\%$; the precent changes are calculated
relative to the baseline period of each model/reanalysis separately
(the time evolution of the different reanalyses is also shown in Supplementary
Fig.~2). In CMIP6 models, a similar intensification by $12\%$ is
only projected to occur by 2080 (with standard deviation of $\pm11$
years across the models), which highlights that climate models may
not only underestimate the recent storms' intensification, but might
also severely under-predict the future intensification of the storms.
Interestingly, the recent weakening of summer storms in the Northern
Hemisphere in reanalyses was also found to occur in climate models
only by the end of the 21st century\cite{Coumou2015}.

The intensification of the storm tracks also suggests a strengthening
in the poleward eddy energy flux (flux by atmospheric perturbations,
such as mid-latitude storms). To examine this we plot in Fig.~\ref{fig:trends}c
the 1979-2018 trends in wintertime poleward transient eddy moist static
energy flux ($v'm'$, where $v$ is meridional velocity, $m$ is moist
static energy and prime denotes eddy terms, Methods) at mid-latitudes;
this energy flux accounts for the poleward eddy flux of heat, moisture
and geopotential (Methods). Similar to the EKE trends, over the last
four decades, $v'm'$ has intensified faster in reanalyses than in
climate models; reanalyses show an intensification in a mean rate
of $6\times10^{12}\,{\rm {TWyr^{-1}}}$, varying between $4.9\times10^{12}\,{\rm {TWyr^{-1}}}$
and $7\cdot10^{12}\,{\rm {TWyr^{-1}}}$ across the reanalyses, while
CMIP6 models show a mean intensification of only $1\times10^{12}\,{\rm {TWyr^{-1}}}$,
where not a single model is able to capture the intensification in
the reanalyses. Decomposing the $v'm'$ trends reveals that the discrepancy
between models and reanalyses is found in the poleward eddy heat and
moisture fluxes, but not in the poleward eddy geopotential flux (Supplementary
Fig.~3). Note that while it is conceivable that the larger trends
in reanalyses might partly stem from larger climatological values,
such direct relation is not found across models, nor across reanalyses
(Supplementary Fig.~4). Moreover, while all reanalyses show larger
EKE and $v'm'$ trends, some reanalyses have similar climatological
EKE and $v'm'$ values as in CMIP6 models.

Similar to the time evolution of EKE, $v'm'$ is also projected to
monotonically intensify over the 21st century (Fig.~\ref{fig:trends}d).
In particular, over the last decade, climate models simulate a $\sim4\%$
increase in $v'm'$ (with standard deviation of $\sim\pm5\%$ across
the models), relative to the 1980-1999 period, while reanalyses show
a $\sim16\%$ increase. A similar $16\%$ increase in climate models
is only projected to occur by 2070 (with standard deviation of $\pm11$
years across the models), which again stresses that climate models
might under-predict the future changes in Southern Hemisphere eddy
energy fluxes, and thus also their large impacts on the distribution
of climate zones in the extra-tropics. Lastly, note that biases in
long-term trends (and climatology\cite{Hodges2017}) might be found
in reanalyses\cite{Chemke2019b}, due to the inclusion of new observed
products and data assimilation methods. However, the use of several
reanalyses, which all show very similar monotonic increase in EKE
and $v'm'$ over the last four decades, together with the fact that
the larger 1979-2018 trends in reanalyses, relative to models, are
evident also on shorter trends (Extended Data Fig.~1), provide us
confidence to use these datasets to evaluate the recent changes in
the storm tracks' activity.

\paragraph{Detection analysis for winter storm tracks intensification}~\\Does the large discrepancy between reanalyses and climate models hinder
the detection of the intensification of winter storm tracks in climate
models? To answer this question we follow previous studies\cite{Hawkins2012,Santer2013a,Chemke2020a}
and analyze the time of emergence of the intensification, out of the
internal climate variability, in both reanalyses and models. This
is done using a signal-to-noise ratio approach, where the signal is
defined as trends of different lengths, all starting from 1979 (the
first year of the reanalyses). The noise is defined as two standard
deviations of all trends, with the same length as the signal, that
arise only due to internal variability (estimated from pre-industrial
control runs, Methods). The time of emergence is defined as the year
when the signal (i.e., the trend) exceeds the noise.

In order to use the signal-to-noise ratio approach we first ensure
that climate models adequately capture the internal variability of
the storm tracks' trends, and thus can be used to assess the 'noise'.
To compare the internal variability of the trends we calculate the
standard deviation across all 10-year, 20-year and 30-year trends
in models and reanalyses over the detrended 1979-2018 period. Similarly,
we calculate the standard deviation across all trends of the same
length from the pre-industrial control runs. We find that reanalyses
and climate models have similar 10-, 20- and 30-year trends variability
(Extended Data Fig.~3). Thus, climate models can be used to evaluate
the internal variability of the storm tracks' trends.

\begin{figure*}
\centering{}\includegraphics[width=1\columnwidth]{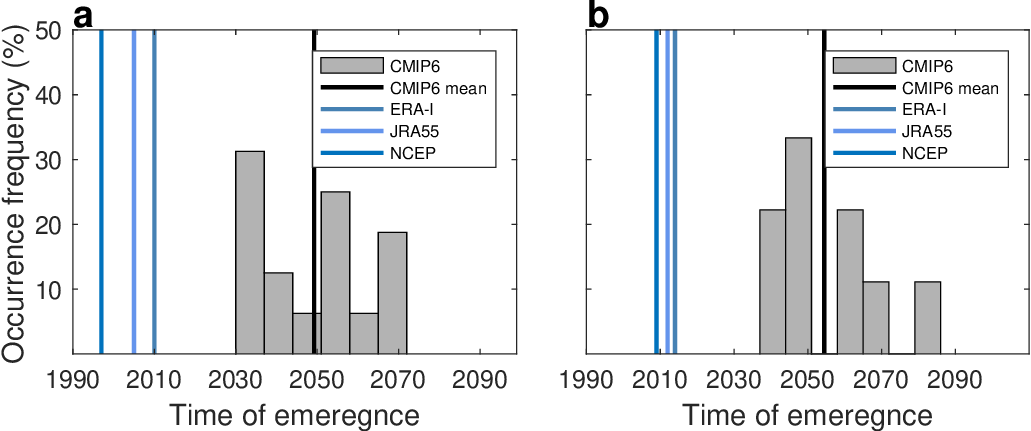}\protect\caption{\label{fig:TOE}\textbf{Time of emergence of mid-latitude storm tracks.}
The occurrence frequency of the time of emergence of \textbf{a,}~EKE
and \textbf{b,}~$v'm'$ across the CMIP6 models (gray). The vertical
blue and black lines show the emergence in each reanalysis (different
shades of blue) and in CMIP6 mean, respectively.}
\end{figure*}

The time of emergence of the intensification of mid-latitude EKE and
$v'm'$ is shown in Fig.~\ref{fig:TOE}. The intensification signal
of EKE in reanalyses emerged from the internal climate variability
in the late '90s-early '00s (Fig.~\ref{fig:TOE}a). In contrast,
the weaker strengthening in climate models significantly delays the
emergence of the signal; the EKE intensification will emerge around
2050 in the multi-model mean, with standard deviation of $\pm14$
years. Similarly, the strengthening of $v'm'$ in reanalyses emerged
around 2010 (Fig.~\ref{fig:TOE}b), while the emergence in climate
models is projected to occur only around 2055 in the multi-model mean
with standard deviation of $\pm14$ years. This highlights that, by
underestimating the magnitude of the storm tracks' changes, climate
models may also underestimate the timing of executing adaption and
mitigation strategies in Southern Hemisphere mid-latitudes.

\paragraph{The source of the models-reanalyses discrepancy}~\\Next, we examine three possible sources of the larger intensification
of mid-latitude storm tracks in reanalyses, relative to models. First,
the evolution of EKE and $v'm'$ in reanalyses is affected by both
their response to external forcings and their internal variability\cite{Hawkins2009,Deser2012,Deser2012a,Thompson2015}.
We thus start by examining whether the size of the CMIP6 ensemble
analyzed here (i.e., 16 model trends) might not be large enough to
capture the internal variability of 40-year trends, and thus also
the trends in reanalyses (i.e., whether internal variability is mostly
responsible for the strong trends in reanalyses). To evaluate the
internal variability in 40-year EKE and $v'm'$ trends we calculate
two standard deviations of all 40-year trends from the pre-industrial
control runs\cite{Thompson2015} (total of 1961 and 1361 EKE and $v'm'$
trends, respectively), and center it around the multi-model mean 1979-2018
trend (red error bars on top of gray bars in Fig.~\ref{fig:trends}a
and \ref{fig:trends}c). While increasing the ensemble size does increase
the range of CMIP6 EKE trends (compare red bars and black dots), even
with a significantly larger ensemble size CMIP6 models still do not
capture the recent trends in reanalyses (the overlap between the uncertainty
in reanalyses $v'm'$ trends and the variability in the model trends
suggests that internal variability could reduce, but not explain,
the models' $v'm'$ bias\cite{Deser2012a,Thompson2015}; compare black
and red error bars). Similarly, increasing the CMIP6 ensemble size
by pooling together all 40-year trends over the 1969-2028 period (total
of 320 trends), or examining the 1979-2018 trends using the Community
Earth System Model (CESM) 40-member ensemble\cite{Kay2015}, where
the spread across its members represents the internal variability
(Methods), yields in similar results of smaller trends in models,
relative to reanalyses (Supplementary Figs.~5-6). Thus, from the
above analyses we conclude that the size of the CMIP6 ensemble is
not likely to explain the smaller trends in models, relative to reanalyses.

Second, the larger EKE and $v'm'$ trends in reanalyses, relative
to models, might stem from the inability of climate models to simulate
the recent cooling of the Southern Ocean surface\cite{Jones2016}.
This model bias, was found to explain the models' inability to capture
the recent trends in annual mean eddy heat fluxes\cite{Chemke2020a}.
To evaluate the effect of recent observed changes in sea surface temperature
(SST) in the models-reanalyses discrepancy we calculate the EKE and
$v'm'$ trends using the Atmosphere Model Intercomparison Project
Phase 6 (AMIP6) runs; in these runs there is no active ocean and sea-ice,
and the SST and sea-ice are prescribed to observations (Methods).
Although we find that correcting the simulated surface changes leads
to slightly larger intensification in models, the intensification
is relatively modest and we conclude that even AMIP6 runs do not capture
the recent trends in reanalyses (green bars in Fig.~\ref{fig:trends}a
and \ref{fig:trends}c). This emphasizes that the inability of climate
models to capture the recent wintertime EKE and $v'm'$ trends does
not stem from biases in simulating SST trends. Interestingly, larger
EKE trends, relative to AMIP6, are also found in NOAA-CIRES-DOE reanalysis
(Supplementary Fig.~7), which, similar to AMIP6, uses observed SST
and sea-ice, but, unlike AMIP6, it (only) assimilates surface pressure.
This suggests, from geostrophic balance, that biases in the structure
of the wind may affect the smaller EKE trends in models, relative
to reanalyses (as further shown below).

Third, since mid-latitude eddies arise from hydrodynamic instability,
we follow previous studies\cite{Chemke2020c}, and further investigate
the source of the models-reanalyses discrepancy by conducting a linear
normal-mode instability analysis. Such analysis allows one to examine
the (maximum) growth rate of mid-latitude eddies for a given mean
atmospheric conditions, which represents the extraction of energy
from the mean flow by the eddies (Methods). In particular, we examine
how changes in the vertical and meridional structures of the mean
flow might explain the recent storm tracks changes, by modulating
the eddy growth rate.

\begin{figure*}
\centering{}\includegraphics[width=1\columnwidth]{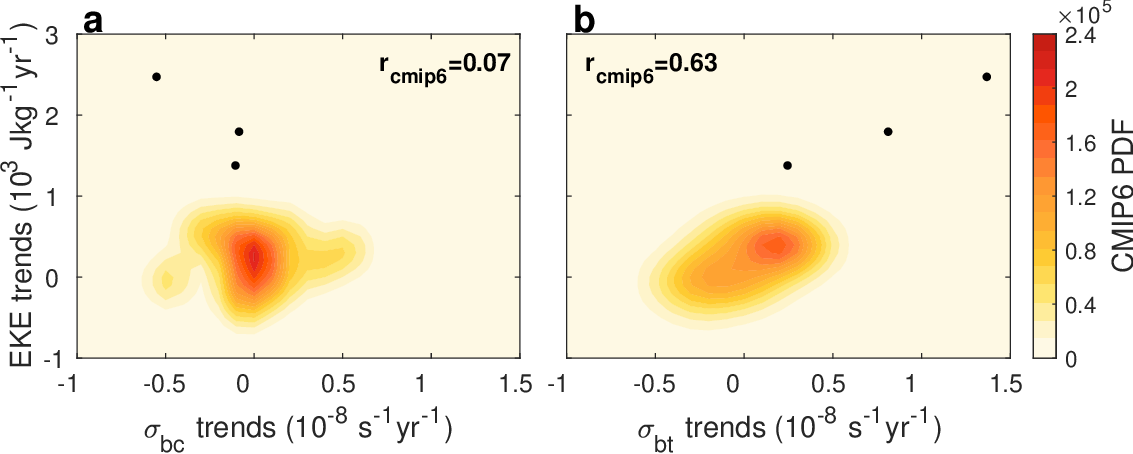}\protect\caption{\label{fig:growth}\textbf{Linear normal-mode instability analysis.}
The 1979-2018 EKE trends plotted against the trends in growth rates
of mid-latitude eddies, calculated from a linear normal-mode instability
analysis, using the \textbf{a,}~vertical (baroclinic, $\sigma_{{\rm {bc}}}$)
and \textbf{b,}~meridional (barotropic, $\sigma_{{\rm {bt}}}$) structure
of the mean flow. Contours show the two-dimensional probability density
function of the trends in CMIP6 models, estimated by fitting a kernel
distribution (the area under each distribution is one), and the correlation
appears in each panel. The black circles show the trends from reanalyses.}
\end{figure*}

To examine the effect of changes in the vertical structure of the
flow (i.e., baroclinicity), we conduct, at each year, a one-dimensional
vertical linear normal-mode instability analysis to the quasi-geostrophic
equations, using the mean zonal wind, static stability and tropopause
height from each reanalysis and each model (Methods). Fig.~\ref{fig:growth}a
shows the two-dimensional probability density function of the CMIP6
1979-2018 EKE trends and the resulting growth rate trends ($\sigma_{{\rm {bc}}}$),
along with the corresponding trends in reanalyses (black dots). The
EKE trends have low correlation with the growth rate trends across
models ($r=0.07$), and reanalyses not only show negative growth rate
trends (which are inconsistent with the EKE trends, and stem from
a reduced stability by the vertically dependent zonal flow and static
stability), but also, in contrast to the EKE trends, reanalyses do
not show larger $\sigma_{{\rm {bc}}}$ trends than the models. Examining
simpler metrics for baroclinicity, such as the Eady growth rate or
mean available potential energy (Methods), as was done in previous
studies\cite{Yin2005,Wu2010,O'Gorman2010b,Harvey2014,Lehmann2014},
yields similar results (Supplementary Fig.~8).

On the other hand, examining the link between EKE and the meridional
structure of the mean flow (i.e., variations in the barotropic component
of the flow), by conducting, using the tropospheric averaged mean
zonal wind ($\overline{u}$), a one-dimensional meridional linear
normal-mode instability analysis to the absolute vorticity equation
(Methods), reveals that the EKE trends are correlated with the growth
rate trends ($\sigma_{{\rm {bt}}}$) across both models ($r=0.63$)
and reanalyses (Fig.~\ref{fig:growth}b). Furthermore, similar to
the EKE trends, reanalyses also show larger $\sigma_{{\rm {bt}}}$
trends than the models. Similar results are found for the components
of $v'm'$ (eddy heat and moisture fluxes, Supplementary Fig.~9).
This suggests that although baroclinicity drives the formation of
mid-latitude storms, here the changes in the storm tracks over recent
decades, and the models-reanalyses discrepancy, seem to stem from
the barotropic part of the flow. Note, that changes in the meridional
structure of the flow might result in the models-reanalyses discrepancy
by not only affecting the barotropic growth of mid-latitude eddies,
but also their baroclinic growth via, for example, the barotropic
governor\cite{James1987}, and by affecting the barotropic decay phase
in eddy lifecycles\cite{Simmons1978}.

To further demonstrate the different meridional structures of the
wind in models and reanalyses, we next examine the 1979-2018 trends
of the second meridional derivative of the mean zonal wind, $\frac{\partial^{2}\overline{u}}{\partial y^{2}}$
(Extended Data Fig.~4), which plays an important role in the growth
of eddies\cite{Vallis2006a} (Methods). For example, in barotropic
instability, positive values of $\frac{\partial^{2}\overline{u}}{\partial y^{2}}$,
which occur over the flanks of the zonal wind (the source regions
for the instability), allows the necessary condition for instability
to be met\cite{Vallis2006a}. Indeed, reanalyses exhibit an increase
in $\frac{\partial^{2}\overline{u}}{\partial y^{2}}$ over the equatorward
and far-end poleward flanks of the wind, which might explain the increased
growth rate in recent decades. In addition, an increase in eddy generation
over the flanks of the jet also suggests convergence of momentum by
the eddies ($-\frac{\partial\overline{u'v'}}{\partial y}$) over the
flanks of the jet, that results in a double peak of $-\frac{\partial\overline{u'v'}}{\partial y}$\cite{Pedlosky1987,Kidston2010}.
Such double peak behavior is also clearly evident in $-\frac{\partial\overline{u'v'}}{\partial y}$
trends in reanalyses (Extended Data Fig.~5). The increase in $\frac{\partial^{2}\overline{u}}{\partial y^{2}}$
over the flanks of the jets and the double peak behavior of $-\frac{\partial\overline{u'v'}}{\partial y}$
are not captured in most CMIP6 models (Extended Data Figs.~4-5),
which suggests that the effect of changes in recent decades in the
meridional structure of the mean zonal wind on the eddy flow are smaller
in climate models, relative to reanalyses, which might lead to the
weaker EKE and $v'm'$ intensification. 

Lastly, previous studies found model biases in the climatological
position of the zonal wind\cite{Kidston2010c,Simpson2016} (which
still exist, but have been improved, in CMIP6 models\cite{Bracegirdle2020})
and were argued to affect the future meridional structure of the wind
(i.e., the shift of the zonal flow). Thus, it is conceivable that
such biases might also affect the model bias in EKE trends. However,
third of the CMIP6 models used here show a poleward bias in the jet's
latitude, relative to reanalyses, while the other two thirds an equatorward
bias (Supplementary Fig.~10; while a larger set of models might result
in a larger bias in the jet's position, still models show both poleward
and equatorward biases\cite{Bracegirdle2020}). Thus, since all models
exhibit weaker EKE trends, relative to reanalyses, the different zonal
wind positions are likely not the source of the EKE discrepancy; reanalyses
and models exhibit different relations between jet latitude and EKE
trends. 

\paragraph{Discussion and conclusions}~\\Using multiple reanalyses, we find that mid-latitude storm tracks,
including their associated poleward energy flux, have substantially
intensified over recent decades in response to external forcing. Climate
models, on the other hand, are found here to significantly underestimate
this intensification, which is only projected to occur in climate
models by the late 21st century. The inability of climate models to
adequately capture the storm tracks intensification, which delays
the detection of the intensification in models by several decades,
questions the skill of climate models to accurately assess the future
changes in the Southern Hemisphere extra-tropics; mid-latitude storm
tracks affect the distribution of heat, precipitation and weather
events (including extreme events) from low subtropical regions to
the high polar regions. We reveal that the biases in climate models
likely arise from biases in the meridional structure of the zonal
flow, and not from a misrepresentation of internal variability in
climate models nor from the models' inability to simulate the recent
cooling in the Southern Ocean. Our analysis highlights the importance
of further investigating observation-based data to assess both the
impacts of human activity on mid-latitude climate, and the limitations
in climate models (especially the biased zonal wind changes) to form
accurate climate-change adaption and mitigation strategies for the
Southern Hemisphere mid-latitudes. 

\pagebreak{}

\begin{addendum}
\item[Acknowledgements:] R.C. is grateful to the WIS support of young scientists, and for Alexander Novoselsky for downloading the data.
\item[Author Contributions:] R.C. analyzed the data and together with Y.M. and J.Y. discussed and wrote the paper.
\item[Competing Interests:] The authors declare that they have no competing interests.
\end{addendum}

\pagebreak{}

\section*{{\normalsize{}References}}
\vspace*{0.4cm}


\pagebreak{}

\section*{{\normalsize{}Methods}}
\vspace*{0.2cm}

\sloppy
\noindent \textbf{EKE} ~\\The Southern Hemisphere wintertime storm tracks' intensity is defined,
following previous studies\cite{Yin2005,O'Gorman2010b,Wu2010,Lehmann2014,Coumou2015},
as the column integrated June-August (JJA) transient eddy kinetic
energy, ${\rm {EKE}}=\frac{1}{g}\intop_{0}^{p_{s}}u'^{2}+v'^{2}dp,$
where $g$ is gravity, $p_{s}$ surface pressure, $u$ and $v$ are
the zonal and meridional winds, respectively, $p$ is pressure and
prime denotes eddy terms, calculated using a bandpass filter of 2-6
days. Using a different bandpass filter to define the transient eddies
(e.g., 3-10 days) yields similar results (Supplementary Fig.~11).
We analyze here the column integrated mid-latitude mean (i.e., averaged
zonally and over $40^{\circ}{\rm {S}-}70^{\circ}{\rm {S}}$) EKE (as
was done in previous studies\cite{O'Gorman2010b,Chang2012,Coumou2015,Chemke2020c})
as the different intensification in models and reanalyses appears
throughout the troposphere (Supplementary Fig.~12) and across most
of the mid-latitudes (Supplementary Fig.~13).

\noindent \textbf{Eddy moist static energy flux} ~\\The poleward transient eddy moist static energy flux is defined as
$v'm'=\frac{2\pi a\cos\phi}{g}\intop_{0}^{p_{s}}c_{p}\overline{v'T'}+L_{v}\overline{v'q'}+\overline{v'\Phi'}dp$,
where $v'T'$, $v'q'$ and $v'\Phi'$ are the meridional eddy heat,
moisture and geopotential fluxes, respectively, $T$ is temperature,
$q$ is specific humidity, $\Phi$ is geopotential, overbar denotes
zonal mean, $c_{p}=1004\,{\rm {Jkg^{-1}K^{-1}}}$ is specific heat
capacity and $L_{v}=2.5\cdot10^{6}\,{\rm {Jkg^{-1}}}$ is latent heat
of vaporization.

\noindent \textbf{CMIP6 models}~\\We use daily and monthly output from 16 models that participate
in the Coupled Model Intercomparison Project Phase 6\cite{Eyring2016}
(CMIP6, Supplementary Table~1; we use all models that have available
daily data for the analysis), and in order to weigh all models equally,
we select only the 'r1i1p1f1' member in four experiments: historical
(through 2014), future scenario SSP5-8.5 (through 2100), historical
with prescribed SST and sea-ice (AMIP6, through 2014), and pre-industrial
control run (with constant 1850 forcings). The constant external forcings
in the control run allows one to evaluate the internal climate variability.
Following previous studies\cite{Santer2013a}, before assessing the
internal climate variability, we first concatenate the detrended last
200 years of each model's pre-industrial run, which yields $\sim2,000$
years of control data. Lastly, we here use the long pre-industrial
runs to assess the internal climate variability, rather than reanalyses
data, since the length of the control runs not only allows one to
gather enough statistics to adequately evaluate the internal variability,
but also to account for the variability of 40-year trends, which cannot
be estimated in reanalyses (as there is only one trend over the 1979-2018
period).

\noindent \textbf{Reanalyses}~\\Reanalyses provide the best approximation for the state of the atmosphere,
as they assimilate air and surface observations in general circulation
models, and thus are used here to examine the recent changes in mid-latitude
storm tracks. Four reanalyses are examined in this study including
JRA-55\cite{Kobayashi2015}, NCEP2\cite{ncep2}, Era-Interim\cite{Dee2011}
and NOAA-CIRES-DOE\cite{Slivinski2019}. We here analyze the Era-Interim,
rather than ERA5\cite{Hersbach2020}, for consistency with the large
body of work done using Era-Interim; nevertheless, similar results
are also evident in ERA5 (Supplementary Fig.~14). Unlike the other
reanalyses, the NOAA-CIRES-DOE reanalysis only assimilates surface
pressure, and uses observed SST and sea-ice. We thus analyze the NOAA-CIRES-DOE
in comparison with AMIP6 runs. We use 6-hourly data from JRA-55, NCEP2
and ERA5, and daily data from Era-Interim and NOAA-CIRES-DOE of zonal
and meridional winds, temperature, specific humidity and geopotential
over the 1979-2018 period; the NOAA-CIRES-DOE (20CRv3.MO) is available
over the 1981-2015 period. 

\noindent \textbf{CESM}~\\We also make use of the Community Earth System Model (CESM)
large ensemble\cite{Kay2015}, which comprises 40 members running
from 1920-2100 under the historical (through 2005) and RCP8.5 (through
2100) forcings. Each member in the ensemble is initialized with a
slightly different air temperature ($\mathcal{O}10^{-14}$), yielding
distinct time evolution over the 20th and 21st centuries. Thus, the
spread across the members represents the internal variability, while
the ensemble mean the forced response to external forcings. Since
daily data is not available from CESM for our analysis, we use monthly
data of kinetic energy. Thus, the EKE is here defined as deviations
from monthly mean.

\noindent \textbf{Linear normal-mode instability analysis}~\\For calculating the growth rates of mid-latitude eddies we follow
previous studies\cite{Chemke2015b,Chemke2016a,Chemke2016b,Chemke2016c,Chemke2017a,Chemke2020c}
and conduct a linear normal-mode instability analysis\cite{growthrate}. For the growth
rate that arise from the vertical structure of the flow (baroclinicity)
we use the quasigeostrophic equations (simplified set of equations
for the mid-latitude flow that include conservation of vorticity in
the interior and of buoyancy in the vertical boundaries) linearized
about a zonal mean state (represented by an overbar), which can be
written (for simplicity in Cartesian coordinates) as follows,

\begin{gather}
\frac{\partial q'}{\partial t}+\overline{u}\frac{\partial q'}{\partial x}+\frac{\partial\psi'}{\partial x}\frac{\partial\overline{q}}{\partial y}=0\,,H_{p}<p<p_{s}\nonumber \\
\frac{\partial}{\partial t}\frac{\partial\psi'}{\partial p}+\overline{u}\text{\ensuremath{\frac{\partial}{\partial x}}}\frac{\partial\psi'}{\partial p}-\frac{\partial\psi'}{\partial x}\frac{\partial\overline{u}}{\partial p}=0,\,p=H_{p},p_{s},\label{eq:qgpv}
\end{gather}
where $q'=\nabla^{2}\psi'+\Gamma\psi'$ is the eddy quasigeostrophic
potential vorticity, $u'=-\frac{\partial\psi'}{\partial y}$ and $v'=\frac{\partial\psi'}{\partial x}$,
$\Gamma=\frac{\partial}{\partial p}\frac{f^{2}}{S^{2}}\frac{\partial}{\partial p}$,
$S^{2}=-\frac{1}{\overline{\rho}\overline{\theta}}\frac{\partial\overline{\theta}}{\partial p}$
is static stability, $\theta$ is potential temperature, $\rho$ is
the density, $\frac{\partial\overline{q}}{\partial y}=\beta-\Gamma\overline{u}$
is the mean quasigeostrophic potential vorticity meridional gradient
(the instability analysis is conducted locally for the mid-latitudes
and hence the mean flow is only vertically dependent), $\beta$ is
the meridional derivative of the Coriolis parameter ($f$) and $H_{p}$
is the tropopause height (defined, following the WMO, as the lowest
level where the vertical temperature gradient crosses the $2{\rm \,{Kkm^{-1}}}$
value).

Eq.~\ref{eq:qgpv} can be written in the form of an eigenvalue problem
by substituting a plane-wave solution, $\psi'={\rm {Re}}{\hat{\psi'}(p)e^{i(kx-\omega t)}},$
where $\hat{\psi'}$ represents the normal modes (the eigenvectors),
$k$ is the zonal wavenumber and $\omega$ is frequency (the eigenvalues).
The resulting vertical eigenvalue problem is then solved for each
year using the vertically dependent wintertime zonal mean mid-latitude
fields (velocity, temperature, tropopause height), and we analyze
the resulting fastest growth rate each year.

For the growth rate that arise from the meridional structure of the
flow (barotropic) we use the linearized absolute vorticity ($\eta=\nabla^{2}\psi+f$)
equation for two-dimensional flow, 
\begin{equation}
\frac{\partial\nabla^{2}\psi'}{\partial t}+\overline{u}\frac{\partial\nabla^{2}\psi'}{\partial x}+\frac{\partial\psi'}{\partial x}\frac{\partial\overline{\eta}}{\partial y}=0,\label{eq:barot}
\end{equation}
where $\frac{\partial\overline{\eta}}{\partial y}=\beta-\frac{\partial^{2}\overline{u}}{\partial y^{2}}$,
and $\psi'=0$ at the meridional boundaries. A plane-wave solution
of the form, $\psi'={\rm {Re}}{\hat{\psi'}(y)e^{i(kx-\omega t)}},$
transforms Eq.~\ref{eq:barot} to an eigenvalue problem, only here,
unlike in Eq.~\ref{eq:qgpv}, the normal modes are a function of
latitude. The latitudinally dependent wintertime vertically averaged
(between $850-300\,{\rm {mb}}$) zonal mean winds are used to solve
the resulting eigenvalue problem each year.

\noindent \textbf{Eady growth rate and MAPE}~\\The Eady growth rate\cite{Vallis2006a} is calculated as $\sigma_{{\rm {Eady}}}=\frac{f\frac{\partial\overline{u}}{\partial z}}{N}$,
where $\frac{\partial\overline{u}}{\partial z}$ is the mean zonal
wind shear and $N^{2}=\frac{g}{\overline{\theta}}\frac{\partial\overline{\theta}}{\partial z}$
is static stability. The Eady growth rate is averaged over the mid-latitudes
and the extratropical troposphere ($850-300\,{\rm {mb}}$). The mean
available potential energy (MAPE) is calculated as $\frac{c_{p}}{2g}\int\gamma\left(\overline{T}^{2}-\widetilde{T}^{2}\right)dp$\cite{Peixoto1992},
where $\gamma=\frac{-\kappa\theta}{pT}\left(\frac{\partial\widetilde{\theta}}{\partial p}\right)^{-1}$
and tilde represents a mean over the mid-latitudes at constant pressure.
MAPE is integrated over the mid-latitude troposphere.

\pagebreak{}

\begin{addendum}
\item[Data Availability:] The data used in the manuscript is publicly
available for CMIP6 data (\url{https://esgf-node.llnl.gov/projects/cmip6/}),
NCEP (\url{https://psl.noaa.gov/}), JRA55 (\url{https://rda.ucar.edu/}, ERA-I and ERA5 (\url{https://www.ecmwf.int}), NOAA-CIRES-DOE (\url{https://www.psl.noaa.gov}) and CESM (\url{https://www.cesm.ucar.edu/projects/community-projects/LENS/data-sets.html}).
\item[Code Availability:] Any codes used in the manuscript are available at \url{https://doi.org/10.5281/zenodo.6434217}.
\end{addendum}

\pagebreak{}

\paragraph{Methods references}~\\

\pagebreak{}

\pagebreak{}

\pagebreak{}

\end{document}